\documentstyle[12pt,hpatex]{article}

\begin{document}

\title{Mapping Between Antisymmetric Tensor \\[2mm]
And Weinberg Formulations\\}

\authors{Valeri V. Dvoeglazov\footnote{On leave of absence from
{\it Dept. Theor. \& Nucl. Phys., Saratov State University,
Astrakhanskaya ul., 83, Saratov\, RUSSIA.}\,
Internet address: dvoeglazov@main1.jinr.dubna.su}}

\address{Escuela de F\'{\i}sica, Universidad Aut\'onoma de Zacatecas \\
Antonio Doval\'{\i} Jaime\, s/n, Zacatecas 98068,
Zac., M\'exico\\ Internet address:  VALERI@CANTERA.REDUAZ.MX
}

\abstract{In the framework of the classical field theory a mapping
between antisymmetric tensor matter fields and Weinberg's $2(2j+1)$
component ``bispinor" fields is considered.  It is shown that such a
mapping exists and equations which describe the $j=1$ antisymmetric tensor
field coincide with the Hammer-Tucker equations completely and with the
Weinberg ones within a subsidiary condition, the Klein-Gordon equation. A
new Lagrangian for the Weinberg theory is proposed.  It is scalar,
Hermitian and contains only the first-order time derivatives of the
fields. The remarkable feature of this Lagrangian is the presence of dual
field functions, considered as parts of a parity doublet.  I study then
origins of  appearance of the dual solutions in the Weinberg equations on
the basis of spinorial analysis and point out the topics which have to be
explained in the framework of a secondary quantization scheme.}

In the present paper the connection between the Weinberg's
$2(2j+1)$- component formulation~\cite{Weinberg,Sankar,Tucker}
and the antisymmetric tensor matter field
description~\cite{Gursey,Hayashi,Kalb,Avd} is studied.
The reason is the elaboration of the Bargmann-Wightman-Wigner-type
quantum  field theory, undertaken by Ahluwalia {\it et al.}~\cite{Ahl2}
(see also~\cite{Ahl}), which brought some hopes on recreation and  further
development of  the Weinberg's $2(2j+1)$ theory.  Finding the likely
relations between various formulations for $j=1$ (and higher spin)
particles could provide a necessary basis for both
practical phenomenological calculations~\cite{Lukaz,Dvoegl2}
and experiment.

I start from the Proca equations for a $j=1$ massive particle
\begin{eqnarray}\label{eq:01}
&&\partial_\mu F_{\mu\nu} = m^2 A_\nu \quad, \\
&&F_{\mu\nu} = \partial_\mu A_\nu - \partial_\nu A_\mu
\end{eqnarray}
in the form given by~\cite{Sankar,Lurie}.
The Euclidean metric,
$x_\mu =  (\vec x, x_4 =it)$ and notation
$\partial_\mu = (\vec \nabla, -i\partial/\partial t)$,
$\partial_\mu^2 = \vec \nabla^{\,2} -\partial_t^2$, are
used. By means of the choice of  $F_{\mu\nu}$ components
as the ``physical" variables I re-write the set of equations as
\begin{equation}\label{eq:eq}
m^2 F_{\mu\nu} =\partial_\mu \partial_\alpha F_{\alpha\nu}
-\partial_\nu \partial_\alpha F_{\alpha\mu}
\end{equation}
and
\begin{equation}\label{eq:2}
\partial_\lambda^2 F_{\mu\nu} = m^2 F_{\mu\nu}\quad.
\end{equation}
It is easy to show that they can be represented in the form
($F_{44}=0$, $F_{4i} =i E_i$ and $F_{jk} =\epsilon_{jki} B_i$;\,
$p_\alpha=-i\partial_\alpha$):
\begin{eqnarray}\label{eq:aux1}
\cases{(m^2 +p_4^2) E_i +p_i p_j E_j +
i\epsilon_{ijk} p_4 p_j B_k=0& \cr
&\cr
(m^2 +\vec p^{\,2}) B_i -p_i p_j B_j +
i\epsilon_{ijk} p_4 p_j E_k =0\quad, &}
\end{eqnarray}
or
\begin{eqnarray}\label{eq:aux}
\cases{\left [ m^2 +p_4^2 +\vec p^{\,2} -
(\vec J \vec p)^2 \right ]_{ij}
E_j +p_4 (\vec J \vec p)_{ij} B_j = 0&\cr
\left [ m^2 +(\vec J \vec p)^2 \right ]_{ij} B_j +
p_4 (\vec J \vec p)_{ij} E_j =0\quad. &}
\end{eqnarray}
Adding and subtracting the  obtained equations yield
\begin{eqnarray}
\cases{m^2 (\vec E +i\vec B)_i + p_\alpha p_\alpha \vec E_i
- (\vec J \vec p)^2_{ij} (\vec E -i \vec B)_j
+ p_4 (\vec J \vec p)_{ij} (\vec B +i\vec E)_j = 0 &\cr
&\cr
m^2 (\vec E - i\vec B)_i + p_\alpha p_\alpha \vec E_i
- (\vec J\vec p)^2_{ij}
(\vec E +i\vec B)_j + p_4 (\vec J \vec p)_{ij}
(\vec B -i\vec E)_j = 0\quad, &}
\end{eqnarray}
where $(\vec J_i)_{jk} = -i\epsilon_{ijk}$ are
the $j=1$ spin matrices.
Equations are equivalent (within a factor $1/2$) to
the Hammer-Tucker equation~\cite{Tucker},
see also~\cite{Dvoegl2,Vlasov}:
\begin{equation}\label{eq:Tucker}
(\gamma_{\alpha\beta}p_\alpha p_\beta
+p_\alpha p_\alpha +2 m^2 ) \psi_1 =0 \quad ,
\end{equation}
in the case of a choice $\chi= \vec E +i\vec B$
and $\varphi =\vec E -i\vec B$,\,\,\,
$\psi_1 = \mbox{column} (\chi , \quad \varphi)$. Matrices
$\gamma_{\alpha\beta}$ are the
covariantly defined matrices of Barut,
Muzinich and Williams~\cite{Barut}.
One can check that the equation (\ref{eq:Tucker}) is
characterized by solutions with a physical dispersion only,
but some points concerned with massless limit should be
clarified properly.

Following to the analysis of ref.~[9b, p.1972],
one can conclude that other equations
with the correct physical dispersion could be
obtained from
\begin{equation}
(\gamma_{\alpha\beta}p_\alpha p_\beta
+a p_\alpha p_\alpha +bm^2) \psi =0 \quad .
\end{equation}
As a result of taking into account $E^2 -\vec p^{\,2} = m^2$ we
draw the conclusion that there exists an infinity number of appropriate
equations provided that $b$ and $a$ are connected as follows:
$$\frac{b}{a+1}=1 \qquad \mbox{or} \qquad \frac{b}{a-1}=1\quad.$$
However, there are only two equations that do not have acausal tachyonic
solutions. The second one (with $a=-1$ and $b=-2$) is
\begin{equation}
(\gamma_{\alpha\beta}p_\alpha p_\beta - p_\alpha p_\alpha - 2m^2)
\psi_2 =0 \quad .
\end{equation}
Thus, we found the ``double" of the Hammer-Tucker equation.
In the tensor form it leads to the equations
dual to (\ref{eq:aux1}):
\begin{eqnarray}
\cases{(m^2 +\vec p^{\,2})C_i - p_i p_j C_j - i\epsilon_{ijk}
p_4 p_j D_k = 0 & \cr
&\cr
(m^2 +p_4^2 )D_i + p_i p_j D_j
- i\epsilon_{ijk} p_4 p_j C_k =0\quad, &}
\end{eqnarray}
which could be re-written in the form, {\it cf.} (\ref{eq:eq}),
\begin{equation}\label{eq:eqd}
m^2 \tilde F_{\mu\nu} =\partial_\mu \partial_\alpha \tilde
F_{\alpha\nu}
-\partial_\nu \partial_\alpha \tilde F_{\alpha\mu} \quad ,
\end{equation}
with $\tilde F_{4i} = iD_i$ and $\tilde F_{jk} =
- \epsilon_{jki} C_i$.
\, \, $C_i$ is an analog of $E_i$ and $D_i$ is an analog of $B_i$\,;\,
$\tilde F_{\mu\nu} ={1\over 2} \epsilon_{\mu\nu\rho\sigma}
F_{\rho\sigma}$, $\epsilon_{1234} = -i$.
I have used above the following
properties of the antisymmetric Levi-Civita tensor
$$ \epsilon_{ijk}
\epsilon_{ijl} =2\delta_{kl}\quad, \quad \epsilon_{ijk} \epsilon_{ilm} =
(\delta_{jl} \delta_{km} -\delta_{jm} \delta_{kl} )\quad,$$
and
$$ \epsilon_{ijk}
\epsilon_{lmn} =  \mbox{Det}\, \pmatrix{\delta_{il} & \delta_{im} &
\delta_{in}\cr
\delta_{jl} & \delta_{jm} & \delta_{jn}\cr
\delta_{kl} & \delta_{km} & \delta_{kn}} \quad. $$

Comparing the structure of the Weinberg equation ($a=0$, $b=1$)
with the Hammer-Tucker ``doubles" one can convince ourselves
that the former can be represented in the tensor form:
\begin{equation}\label{eq:3}
m^2 F_{\mu\nu} =\partial_\mu \partial_\alpha F_{\alpha\nu}
-\partial_\nu \partial_\alpha F_{\alpha\mu} + {1\over 2}
(m^2 - \partial_\lambda^2) F_{\mu\nu}\quad.
\end{equation}
However, as we learnt, it is possible to build a ``double" equation:
\begin{equation}\label{eq:4}
m^2 \tilde F_{\mu\nu} =\partial_\mu \partial_\alpha \tilde
F_{\alpha\nu}
-\partial_\nu \partial_\alpha \tilde F_{\alpha\mu} +
{1\over 2} (m^2 -\partial_\lambda^2) \tilde F_{\mu\nu}\quad.
\end{equation}
Thus, the Weinberg's set of equations could be written in the form:
\begin{eqnarray}\label{eq:a1}
(\gamma_{\alpha\beta} p_\alpha p_\beta + m^2 )\psi_1 &=& 0\quad,\\
\label{eq:a2}
(\gamma_{\alpha\beta} p_\alpha p_\beta - m^2) \psi_2 &=& 0 \quad.
\end{eqnarray}
Thanks to the Klein-Gordon equation (\ref{eq:2}) these equations
are equivalent to the Proca tensor equations (and to
the Hammer-Tucker ones) in a free case.
However, if interaction is included, one cannot say that.
The general solution describing a $j=1$ particle is presented as a
superposition
\begin{equation}\label{eq:super}
\Psi^{(1)} = c_1 \psi_1^{(1)} + c_2 \psi_2^{(1)} \quad,
\end{equation}
where the constants $c_1$
and $c_2$ are to be defined from the boundary, initial
and normalization conditions.

Let me note a surprising fact:
while both the Proca equations (or the Hammer-Tucker ones)
and the Klein-Gordon equation do not possess ``non-physical"
solutions, their sum, Eqs. (\ref{eq:3},\ref{eq:4}) or
the Weinberg equations (\ref{eq:a1},\ref{eq:a2}), acquires tachyonic
solutions. For the following it is also useful to note some
remarkable features of this set of equations.
Equations (\ref{eq:a1}) and (\ref{eq:a2}) could be
re-casted in another form (index $``T"$ denotes a transpose matrix):
\begin{eqnarray}\label{eq:a11}
\left [\gamma_{44} p_4^2 +2\gamma_{4i}^T p_4 p_i +\gamma_{ij}p_i p_j
-m^2\right ] \psi_1^{(2)} &=&0 \quad ,\\ \label{eq:a21}
\left [\gamma_{44} p_4^2 +2\gamma_{4i}^T p_4 p_i +\gamma_{ij}p_i p_j
+m^2 \right ] \psi_2^{(2)} &=&0 \quad ,
\end{eqnarray}
respectively, if understand
$\psi_1^{(2)}=\mbox{column} (B_i + iE_i ,\quad B_i -iE_i)
=i\gamma_5 \gamma_{44} \psi_1^{(1)}$ and
$\psi_2^{(2)} =\mbox{column} (D_i + i
C_i, \quad D_i - iC_i )= i\gamma_5 \gamma_{44} \psi_2^{(1)}$.
The general solution is again a linear combination
\begin{equation}
\Psi^{(2)} =c_1 \psi_1^{(2)} + c_2 \psi_2^{(2)}\quad.
\end{equation}
From, {\it e.g.}, Eq. (\ref{eq:a1}), dividing $\psi^{(1)}_1$
into longitudinal and transversal parts one can come to
the equations:
\begin{eqnarray}
\lefteqn{\left [{\cal E}^2 -\vec p^{\,2}\right ](\vec E +
i\vec B)^{\parallel}
-m^2 (\vec E - i\vec B)^{\parallel} +\nonumber}\\
&+&\left [{\cal E}^2 +\vec p^{\,2}- 2{\cal E} (\vec J\vec p)\right ]
(\vec E + i \vec B)^{\perp} - m^2 (\vec E - i \vec B)^{\perp} =0\quad,
\end{eqnarray}
\begin{eqnarray}
\lefteqn{\left [{\cal E}^2 -\vec p^{\,2}\right ](\vec E - i
\vec B)^{\parallel} -m^2 (\vec E + i \vec B)^{\parallel} +\nonumber}\\
&+&\left [{\cal E}^2 + \vec p^{\,2}+2{\cal E} (\vec J\vec p)\right
] (\vec E - i \vec B)^{\perp} - m^2 (\vec E + i \vec B)^{\perp} =0 \quad.
\end{eqnarray}
Therefore, in classical field theory antisymmetric tensor
matter fields are the fields with the transversal components in massless
limit ({\it cf.} with a quantized case,
ref.~\cite{Hayashi,Kalb,Avd,Dvoegl2} and with the remark in the end of the
paper).\footnote{Let me also mention that the equations (4.21) and (4.22)
of the paper~[1b,p.B888] may be insufficient for describing a $j=1$
massless field as noted absolutely correctly in refs.~[9b], see
also~[11c,e].  As a matter of fact their inadequate application leads to
speculations on  the violation of the Correspondence Principle.
Unfortunately, the author of ref.~[1b] missed the fact that the matrix
$(\vec J\vec p)$ has no the inverse one.}

Under the transformations $\psi_1^{(1)}
\rightarrow \gamma_5 \psi_2^{(1)}$ or $\psi_1^{(2)}
\rightarrow \gamma_5 \psi_2^{(2)}$
the set of equations (\ref{eq:a1}) and (\ref{eq:a2}), or
(\ref{eq:a11}) and (\ref{eq:a21}), are invariant.
The origin of this fact is the dual invariance of the
Proca equations.  In the matrix form
dual transformations correspond to the chiral transformations
(see about these relations, {\it e.g.}, ref.~\cite{Strazhev}).

Another equation has been proposed in refs.~\cite{Sankar,Ahl2}
\begin{equation}\label{eq:Sankar}
(\gamma_{\alpha\beta} p_\alpha p_\beta +
\wp_{u,v} m^2 )\psi =0\quad,
\end{equation}
where $\wp_{u,v}=i(\partial/\partial t)/E$,
what distinguishes
$u$- (positive-energy) and $v$- (negative-energy) solutions.
For instance, in ~[8a,footnote 4] it is claimed that
\begin{equation}
\psi^+_\sigma (x) = \frac{1}{(2\pi)^3} \int
\frac{d^3 p}{2\omega_p} u_\sigma (\vec p) e^{ipx} \quad,
\end{equation}
$\omega_p =\sqrt{m^2 +\vec p^{\,2}}$, $p_\mu x_\mu
=\vec p \vec x -Et$, must be described by the equation
(\ref{eq:a1}), in the meantime,
\begin{equation}
\psi^-_\sigma (x) =
\frac{1}{(2\pi)^3} \int \frac{d^3 p}{2\omega_p}
v_\sigma (\vec p) e^{-ipx}\quad,
\end{equation}
from equation (\ref{eq:a2}).
The analysis of this question
and the comparison of the models based on the set of
equations obtained here and in ref.~\cite{Ahl2}
(as well as the discussion of consequences
of new constructs in the $(j,0)\oplus (0,j)$ representation
space and their differences from the papers
of~\cite{Weinberg,Sankar,Tucker}) are left for further
publications.

Let me consider the question of the ``double" solutions
on the ground of spinorial analysis. In ref.~\cite[p.1305]{Sankar}
(see also~\cite[p.60-61]{Landau})
relations between the Weinberg bispinor (bivector, indeed)
and symmetrical spinors of $2j$ rank have been discussed.
It was noted there: ``{\it
The wave function may be written in terms of two
three-component functions $\psi=column
(\chi \quad \varphi)$,
that, for the continuous group, transform independently of
each other and that are related to two symmetrical
spinors:}
\begin{eqnarray}
&&\chi_1 = \chi_{\dot 1\dot 1} , \quad \chi_2 = \sqrt{2}
\chi_{\dot 1 \dot 2} , \quad \chi_3 =
\chi_{\dot 2 \dot 2} \quad,\\
&& \varphi_1 = \varphi^{11} , \quad \varphi_2 = \sqrt{2}
\varphi^{12} , \quad \varphi_3 = \varphi^{22} \quad,
\end{eqnarray}
{\it when the standard representation for the spin-one
matrices, with $S_3$ diagonal is used.}"

Under the inversion operation we
have the following rules~\cite[p.59]{Landau}:
$\varphi^\alpha \rightarrow \chi_{\dot\alpha}$,\,
$\chi_{\dot
\alpha} \rightarrow \varphi^{\alpha}$,\,
$\varphi_\alpha \rightarrow -\chi^{\dot \alpha}$
and $\chi^{\dot\alpha}\rightarrow -\varphi_\alpha$.
Hence, we deduce (if understand $\chi_{\dot\alpha\dot\beta}
=\chi_{\{\dot\alpha} \chi_{\dot\beta\}}$\, ,\, $\varphi^{\alpha\beta}
=\varphi^{\{\alpha}\varphi^{\beta\}}$)
\begin{eqnarray}
&&\chi_{\dot 1 \dot 1} \rightarrow \varphi^{11} \quad, \quad
\chi_{\dot 2 \dot 2} \rightarrow \varphi^{22} \quad, \quad
\chi_{ \{ \dot 1 \dot 2 \} } \rightarrow
\varphi^{ \{ 12 \} } \quad,\\
&& \varphi^{11} \rightarrow \chi_{\dot 1 \dot 1} \quad, \quad
\varphi^{22} \rightarrow \chi_{\dot 2 \dot 2} \quad, \quad
\varphi^{ \{ 12 \} } \rightarrow \chi_{ \{ \dot 1 \dot 2 \} } \quad.
\end{eqnarray}
However, this definition of symmetrical spinors
of the second rank $\chi$ and $\varphi$ is ambiguous.
We are also able to define $\tilde \chi_{\dot\alpha\dot\beta}
=\chi_{\{\dot\alpha} H_{\dot\beta\}}$ and
$\tilde \varphi^{\alpha\beta} = \varphi^{\{\alpha}
\Phi^{\beta\}}$,
where $H_{\dot\beta} = \varphi_{\beta}^{*}$,
$\Phi^{\beta} =(\chi^{\dot\beta})^{*}$.
It is easy to show that in the framework
of the second definition we have under the inversion operation:
\begin{eqnarray}
&&\tilde \chi_{\dot 1 \dot 1} \rightarrow
-\tilde \varphi^{11} \quad , \quad
\tilde \chi_{\dot 2 \dot 2} \rightarrow
- \tilde \varphi^{22} \quad , \quad
\tilde \chi_{ \{ \dot 1 \dot 2 \} } \rightarrow
- \tilde \varphi^{ \{ 12 \} } \quad, \\
&&\varphi^{11} \rightarrow -\tilde \chi_{\dot 1 \dot 1}\quad , \quad
\tilde \varphi^{22} \rightarrow -\tilde \chi_{\dot 2
\dot 2}\quad , \quad
\tilde \varphi^{ \{ 12 \} } \rightarrow
-\tilde \chi_{ \{ \dot 1 \dot 2 \} } \quad .
\end{eqnarray}

The Weinberg bispinor
$(\chi_{\dot\alpha\dot\beta} \quad \varphi^{\alpha\beta})$
corresponds to the equations (\ref{eq:a11}) and (\ref{eq:a21}) ,
meanwhile
$(\tilde \chi_{\dot\alpha\dot\beta}\quad
\tilde\varphi^{\alpha\beta})$, to
the equation (\ref{eq:a1}) and (\ref{eq:a2}).

Similar conclusions can be drawn in
the case of the parity definition
as $P^2 = -1$. Transformation rules are then
$\varphi^\alpha \rightarrow i\chi_{\dot\alpha}$, $\chi_{\dot\alpha}
\rightarrow i\varphi^\alpha$,
$\varphi_\alpha \rightarrow -i\chi^{\dot\alpha}$
and $\chi^{\dot\alpha}\rightarrow -i\varphi_\alpha$,
ref.~\cite[p.59]{Landau} .
Hence, $\chi_{\dot\alpha\dot\beta} \leftrightarrow
-\varphi^{\alpha\beta}$
and $\tilde \chi_{\dot\alpha\dot\beta} \leftrightarrow
- \tilde \varphi^{\alpha\beta}$, but $\varphi^{\alpha}_{\quad\beta}
\leftrightarrow \chi_{\dot\alpha}^{\quad\dot\beta}$ and $\tilde
\varphi^{\alpha}_{\quad\beta} \leftrightarrow
\tilde \chi_{\dot\alpha}^{\quad\dot\beta}$\quad.

Next, one can propose, {\it e.g.}, the following  Lagrangian
for physical fields $\psi_1^{(1)}$ and $\psi_2^{(2)}$ chosen:
\begin{equation}\label{eq:Lagran}
{\cal L} = \partial_\mu \psi_1^\dagger \widetilde \gamma_{\mu\nu}
\partial_\nu \psi_2
+\partial_\mu \psi_2^\dagger \gamma_{\mu\nu}
\partial_\nu \psi_1 + m^2  \psi_2^\dagger \psi_1
+ m^2 \psi_1^\dagger \psi_2 \quad.
\end{equation}
It is scalar
(as opposed to ref.~[11c]), Hermitian
({\it cf.} ref.~\cite{Greiner}) and of the
first order in time derivatives of fields
(as opposed to ref.~[8b]).
It is easy to check that $\widetilde \gamma_{\mu\nu}=
\gamma_{44} \gamma_{\mu\nu}\gamma_{44}$
transforms as a tensor, therefore the Lagrangian is
still scalar. It leads to the equations
which have solutions in spite of the fact
that a procedure of adding the Hermitian-conjugated part
in previous attempts led to an inconsistent theory (when
the Euclidean metric was used), ref.~\cite{Greiner}.
I would still like to notice that two dual field
functions are used and they are considered
as independent ones in the present formulation.
The Lagrangian (\ref{eq:Lagran})
leads to the equations of another set of Weinberg ``doubles",
each of those has solution.\footnote{On
first sight the second equation
obtained from this Lagrangian differs from the Weinberg ``double",
Eq. (\ref{eq:a2}). However, let us not forget the discussion
surrounding Eq. (\ref{eq:a21}).}
In the case of the use of a pseudoeuclidean metric (when
$\gamma_{0i}$ is chosen to be anti-Hermitian)
it is possible to write
the Lagrangian following for F. D. Santos and H. Van Dam,
ref.~[18b]:
\begin{equation}
{\cal L} = \partial^\mu \bar \psi \gamma_{\mu\nu} \partial^\nu \psi
-m^2 \bar \psi \psi \quad.
\end{equation}
It does not lead to the difficulties noted by W. Greiner and
H. M. Ruck (since there is no necessity to add
the Hermitian-conjugated part).
However, in the papers of F. Santos and H. Van Dam,
ref.~\cite{Santos},
the possibility of  appearance of the ``doubles"
has not been considered (neither in any other
paper on the $2(2j+1)$ formalism, to my knowledge).

I conclude: Both the theory of Ahluwalia {\it et al.} and the
model based on the use of $\psi_1$ and $\psi_2$
are connected with the antisymmetric tensor matter field
description. The models are both possible at
the classical level. However,
even at the classical level they may lead to different
predictions, which also differ from the previous considerations.
They have to be quantized consistently. Special attention
should be paid to the translational and rotational invariance
(the conservation of energy-momentum
and angular momentum, indeed), the interaction
representation, causality, locality and covariance of theory,
{\it i.e.} to all topics, which are the axioms of the modern
quantum field theory ~\cite{Itzykson,Bogol}.
A consistent theory has also to take into account the degeneracy
of states.\footnote{Namely, two
dual functions $\psi_1$ and $\psi_2$ (or $F_{\mu\nu}$
and $\tilde F_{\mu\nu}$, the ``doubles")
are considered to yield the
same spectrum. Unfortunately, in previous formulations
of the dual theories  many specific
features of such a consideration have not been taken into account.
I am going to present
a detailed version of this Weinberg-type  model in the
forthcoming publications.}

Then, I would like to draw reader's attention at the problem
put forward in~[11c]. The papers~\cite{Hayashi,Kalb,Avd}
have proved that the quantized $F_{\mu\nu}$ tensor field
describes the particles with the longitudinal component
only in the massless limit, the $0^+$ particle. In the meantime, the
quantized $\tilde F_{\mu\nu}$ field describes the $0^-$ massless
particle.\footnote{Not all the authors agreed with this
conclusion, {\it e.g.},~\cite{Gursey,Boyarkin,Ahl2}.}
How is the Weinberg theorem\footnote{Let
me recall that the Weinberg theorem
states: {\it The fields constructed from massless particle operator
\,\,$a (\vec p,\lambda)$\,\, of the definite helicity transform according
to representation\,\, $(A,B)$ \,\,such that $\qquad B-A=\lambda$.}}  to be
treated in this case (we use the  $(1,0) +(0,1)$ representation, however,
$\lambda=0$)? Why do the Weinberg equations seem to describe the
transversal fields in the classical theory and the longitudinal
fields in the quantized theory.  Finally, since this formulation
appears to be related\footnote{In fact, the equations
(\ref{eq:eq},\ref{eq:eqd}) are the ones for the Hertz tensors,
which are connected with the field tensors, see the dual theory of
electrodynamics which uses a formalism with the Hertz tensors,
{\it e.g.}, ref.~[21b]).} to the dual theories let me reproduce some
references which could be useful in further development of the theory.
Dual formulations of electrodynamics and massive vector theory based on
$F_{\mu\nu}$ and $\tilde F_{\mu\nu}$ have been considered in
refs.~\cite{Strazhev,Cabbibo,Boyarkin,Recami,Schwarz}.  The interaction of
the Dirac field with the dual fields $F_{\mu\nu}$ and $\tilde F_{\mu\nu}$
has been considered in ref.~\cite{Lavrov} (what, indeed, implies
existence of the anomalous electric dipole moment of a fermion).  The dual
formulation of the Dirac field has also been considered, {\it e.
g.}, ref.~\cite{Brana} (see also~\cite{Das}).

{\bf Acknowledgments.}
I am glad to express my gratitude to Prof. D. V. Ahluwalia for
reading of a preliminary version of the manuscript.
His answers on my questions were very incentive to
the fastest writing the paper. The firm support
that my thoughts are not groundless was in one of the letters.
I greatly appreciate valuable discussions
with Profs. J. Beckers, A. Mondragon, M.~Moshinsky,
Yu. F. Smirnov and A. Turbiner.

Recently after writing this paper  I learnt about the papers of Prof. M.
W.  Evans published in FPL, FP and Physica B as well as about his several
books.  As a matter of fact they deal (from a different viewpoint) with
the questions put forward in my series. I am grateful to Prof.  Evans for
many private communications on his concept of the longitudinal $B^{(3)}$
field of electromagnetism and I appreciate his encouragements.

I am grateful to Zacatecas University (M\'exico) for a professorship.

\end{document}